\documentclass[12pt]{article}
\usepackage[dvips]{color}
\usepackage{amssymb}
\usepackage{epsfig}

\usepackage[dvips]{color}
\usepackage{epsfig}
\usepackage{amsmath}
\usepackage{graphicx}
\def\be {\begin{equation}}
\def\ee {\end{equation}}
\def\ba {\begin{eqnarray}}
\def\ea {\end{eqnarray}}

\def\bi {\begin{itemize}}
\def\ei {\end{itemize}}
\begin{document}
\begin{large}
\title{\bf  { Quintom dark energy in the DGP braneworld cosmology}}
\end{large}
\author{  P. Moyassari $^1$ \hspace{0.2 cm}and\hspace{0.2 cm}  M. R. Setare$^2$\thanks{%
email: rezakord@ipm.ir}\\
$^1${\small Department of Physics, Nanjing University 22 Hankou
Road,}\\ {\small Nanjing 210093, China.}\\
 $^2${\small Department of Science,
Payame Noor University, Bijar, Iran.}\\} \maketitle
 \begin{abstract}
 In this paper we consider a $Z_2$ symmetrical 3-brane embedded in a 5-dimensional
spacetime. We study the effective Einstein equation and acceleration
condition in presence of the quintom dark energy fluid as the bulk
matter field. It is shown that the time-dependent bulk quintom field
induces a time-dependent cosmological constant on the brane. In the
framework of the DGP model, the effective Einstein equation is
obtained in two different cases: i) where the quintom field is
considered as the bulk matter field and the brane is empty and, ii)
where the quintom dark energy is confined on the brane and the bulk
is empty. We show that in both cases one could obtain a
self-inflationary solution at late time in positive branch
$\epsilon=1$, and an asymptotically static universe in negative
branch $\epsilon=-1$.
 \end{abstract}
 \section{Introduction}
Recent observations of type Ia supernova (SNIa) and WMAP
\cite{exp,wmap} indicate that our universe is currently undergoing
an accelerating expansion, which
 confront the fundamental theories with great challenges and also make the researches on this problem a major endeavor in modern
 astrophysics and cosmology. Missing energy density - with negative pressure - responsible for
this expansion has been dubbed dark energy. Wide range of scenarios
have been proposed to explain this acceleration while most of them
can not explain all the features of universe or they have so many
parameters that makes them difficult to fit. The models which have
been discussed widely in literature are those which consider vacuum
energy (cosmological constant) \cite{cosmo} as dark energy,
introduce fifth elements and dub it quintessence \cite{quint} or
scenarios named phantom \cite{phant} with $w<-1$ , where $w$ is
parameter of state. A challenging issue is that the time-dependent
dark energy gives a better fitting than a cosmological constant, and
in particular the analysis of the properties of dark energy from
recent observations mildly favor models with $w$ crossing -1 at
redshift $z\approx0.2$. Neither the quintessence nor the phantom
alone can fulfill the transition from $w>-1$ to $w<-1$ and vice
versa. Although for k-essence \cite{k-essence} one can have both
$w\ge -1$ and $w<-1$, it has been lately considered by Ref
\cite{Vikman1} that it is very difficult for k-essence to get $w$
across $-1$ during evolving. But one can show
\cite{{quintom},{quint2}} that considering the combination of
quintessence and phantom in a joint model, the transition can be
fulfilled. This model, dubbed quintom, can produce a better fit to
the data than more familiar models with $w\geq-1$. In other words,
the quintom model of dark energy represents a transition of dark
energy equation of state from $w>-1$ to $w<-1$, or vice versa,
namely from $w<-1$ to $w>-1$ is also one realization of quintom, as
can be seen clearly in \cite{ref}. Although the models with negative
kinetic term are often plagued by instability, there are
possibilities that these models might be phenomenologically viable
if considered as effective field theories \cite{stability,Vikman2}.

An alternative way of explaining the observed acceleration of the
late universe is to modify gravity at large scales. A well-studied
model of modified gravity is the Dvali-Gabadadze-Porrati (DGP)
braneworld model \cite{2} where the brane is embedded in the flat
bulk with infinite extra dimension. In this model gravity leaks of
the 4-dimensional brane universe into 5-dimensional bulk spacetime
at large scales. The inclusion of a graviton kinetic term on the
brane recovers the usual gravitational force law scaling, $1/r^2$,
at short distances, but at large distances it asymptotes to the
5-dimension scaling, $1/r^3$. Motivated by string/M theory, the
AdS/CFT correspondence, and the hierarchy problem of particle
physics, braneworld models were studied actively in recent years
\cite{Hora96}-\cite{Rand99}. In these models, our universe is
realized as a boundary of a higher dimensional spacetime. The matter
particles can not freely propagate in those large extra dimensions,
but must be constrained to live on a 4-dimensional submanifold. The
DGP model has a large scale/low energy effect of causing the
expansion rate of the universe to accelerate. In almost all of works
on braneworld models, the 5-dimensional bulk spacetime is assumed to
be vacuum except for the presence of the cosmological constant, and
the matter fields on the brane are regarded as responsible for the
dynamics of the brane. However, from the unified theoretic point of
view, the gravitational action is not necessarily the
Einstein-Hilbert action. In fact, string theory tells us that the
dimensionally reduced effective action includes not only
higher-order curvature terms but also dilatonic gravitational scalar
fields. Thus at the level of the low-energy 5-dimensional theory, it
is naturally expected that there appears a dilaton-like scalar field
in addition to the Einstein-Hilbert action \cite{lu}. Hence it is of
interest to investigate how such a scalar field in the 5-dimensional
theory affects the braneworld \cite{Ell,carol}.

In this paper our main motivation is investigating the effects of
the bulk quintom field on the evolution of the universe in the
braneworld scenario and in the DGP model. We first review the
braneworld scenario in presence of the bulk matter field in section
2. We study the acceleration condition for the universe with quintom
dark energy in the bulk and show that the time-dependent bulk
quintom field alters the brane as a time-dependent cosmological
constant which is related to the quintom potential on the brane. In
section 3 we obtain the generalized Einstein equation in the DGP
model in presence of the tension and the bulk matter field. In the
two following sections we investigate whether it is possible to have
a late time accelerating phase on the brane when there is a quintom
dark energy fluid in the bulk and the brane is empty; or inversely,
when there is a quintom dark energy fluid on the brane and the bulk
is empty.

\section{Effective Einstein equation on the braneworld}
In this section we briefly review the braneworld scenario in
presence of the tension and bulk matter field. In the braneworld
scenario, our 4-dimensional world is described by a domain wall
(brane) in $5$-dimensional spacetime. We consider an ansatz for the
5-dimensional metric of the form
\begin{equation}\label{1}
    ds^2=-n^2(t,y)dt^2+a^2(t,y)\gamma_{ij}dx^idx^j+b^2(t,y)dy^2,
    \end{equation}
where $y$ is the coordinate of the fifth dimension and $\gamma_{ij}$
is a maximally symmetric 3-dimensional metric. We will use $k$ to
parameterize the spatial curvature and assume that the brane is a
hypersurface defined by $y=0$. We shall be interested in the model
described by the action
\begin{equation}\label{1n}
    S=\int d^5x\sqrt{-g}(\frac{1}{2k^2_5}{\hspace{0.3cm}R^{\hspace{-0.7cm}(5)}}{\hspace{0.4cm}-\Lambda+{\mathcal{L}_B^{mat})}+\int
    d^4x\sqrt{-q}(-\xi+\mathcal{L}_b^{mat})},
\end{equation}
where ${\hspace{0.3cm}R^{\hspace{-0.7cm}(5)}}{\hspace{0.4cm}}$ is
the scalar curvature of the 5-dimensional metric $g_{AB}$, $\Lambda$
is the bulk cosmological constant, $\xi$ is the brane tension,
$k^2_5=8\pi G_5$, and $q_{AB}=g_{AB}-n_An_B$ ($n_A $ is the unit
vector normal to the brane and $A,B=0,1,2,3,5$) is the induced
metric on the 3-brane. The 5-dimensional Einstein equation can be
written as
\begin{equation}\label{2}
R^{\hspace{-0.7cm}(5)}_{AB}-\frac{1}{2}g_{AB}{\hspace{0.3cm}R^{\hspace{-0.7cm}(5)}}{\hspace{0.4cm}=k^2_5(-\Lambda
g_{AB}+T_{AB}+S_{\mu\nu}\delta^\mu_A\delta^\nu_B\delta(y_b))},
\end{equation}
here $\delta(y_b)=\frac{\delta(y)}{b}$, $T_{AB}$ is the energy
momentum tensor of the bulk matter and the last term corresponds to
the matter content on the brane
\begin{equation}\label{3}
    S_{\mu\nu}=-\xi g_{\mu\nu}+\tau_{\mu\nu}.
\end{equation}
The non-zero components of the 5-dimensional Einstein equation are

\begin{eqnarray}
  &&3\{-\frac{\dot{a}}{n^2a}(\frac{\dot{a}}{a}+\frac{\dot{b}}{b})+\frac{1}{b^2}(\frac{a''}{a}+\frac{a'}{a}
  (\frac{a'}{a}-\frac{b'}{b}))-\frac{k}{a^2}\} = k^2_5(-\Lambda+T^0_0+S^0_0\delta(y_b)), \label{3n}\\
  &&\frac{1}{b^2}\delta_{j}^i\{\frac{a'}{a}(\frac{a'}{a}+\frac{n'}{n})-
  \frac{b'}{ba^2}(\frac{n'}{n}+2\frac{a'}{a})+\frac{a''}{a}+\frac{n''}{n}\}+
  \nonumber
  \\&&\frac{1}{n^2}\delta_j^i\{\frac{\dot{a}}{a}(-\frac{\dot{a}}{a}+2\frac{\dot{n}}{n})-2\frac{\ddot{a}}{a}+\frac{\dot{b}}{b}
  (-2\frac{\dot{a}}{a}+\frac{\dot{n}}{n})-\frac{\ddot{b}}{b}\}-k\delta_j^i = k^2_5(-\Lambda+T^i_j+S^i_j\delta(y_b)), \label{3nm}\\
  &&3\{\frac{n'}{n}\frac{\dot{a}}{a}+\frac{a'}{a}\frac{\dot{b}}{b}-\frac{\dot{a'}}{a} \}= k^2_5T_{05}, \label{3nb}\\
 && 3\{\frac{a'}{ab^2}(\frac{a'}{a}+\frac{n'}{n})-\frac{1}{n^2}(\frac{\dot{a}}{a}(\frac{\dot{a}}{a}-\frac{\dot{n}}{n})+\frac{\ddot{a}}{a})-\frac{k}{a^2} \}
   = k^2_5(-\Lambda+T^5_5)\label{3nv},
    \end{eqnarray}
where primes indicate derivatives with respect to $y$, while dots
derivatives with respect to $t$. Assuming a perfect fluid on the
brane
\begin{equation}\label{4n}
    \tau^\mu_\nu=diag(-\rho_b,p_b,p_b,p_b),
\end{equation}
and a quintom field in the bulk space containing the normal scalar
field $\phi(t,y)$ and negative kinetic scalar field $\sigma(t,y)$,
with the Lagrangian expressed as the following form
\begin{equation}\label{4a}
    \mathcal{L}^{mat}_B=\frac{1}{2}g^{AB}(\phi_{,A}\phi_{,B}-\sigma_{,A}\sigma_{,B})
    +V(\phi,\sigma).
\end{equation}
 According to this action, the energy momentum
 tensor of the bulk quintom field is given by
\begin{equation}\label{5}
    T_{AB}=\phi_{,A}\phi_{,B}-\sigma_{,A}\sigma_{,B}-g_{AB}(\frac{1}{2}g^{CD}
    (\phi_{,C}\phi_{,D}
    -\sigma_{,C}\sigma_{,D})+V(\phi,\sigma)).
\end{equation}
In order to focus on the cosmological evolution on the brane we use
the Gaussian normal coordinates $\left(b(y,t)=1\right)$
\cite{tetradisbulk1}. Thus the equations of motion of the scalar
field $\phi$ and $\sigma$ in bulk space are
\begin{eqnarray}\label{5n}
  -\ddot{\phi}-(3\frac{\dot{a}}{a}-\frac{\dot{n}}{n})\dot{\phi}+n^2[(\frac{n'}{n}+3\frac{a'}{a})\phi'+\phi'']-n^2\frac{\partial V(\phi,\sigma)}
  {\partial\phi}&=& \frac{\delta\mathcal{L}_b^{mat}}{\delta\phi}\delta(y_b), \nonumber\\
  & &\\
  -\ddot{\sigma}-(3\frac{\dot{a}}{a}-\frac{\dot{n}}{n})\dot{\sigma}+n^2[(\frac{n'}{n}+3\frac{a'}{a})\sigma'+\sigma'']+n^2\frac{\partial V(\phi,\sigma)}{\partial\sigma}&=&
   -\frac{\delta\mathcal{L}_b^{mat}}{\delta\sigma}\delta(y_b).\nonumber
\end{eqnarray}
We are interested in studying the Einstein equation in presence of a
quintom field in the bulk at the location of the brane. Without
losing generality we choose  $n(t,0)=1$ which can be achieved by
scaling the time coordinate. As is well known, the presence of the
brane leads to a singular term proportional to $\delta$-function in
$y$ on the right-hand sides of the Einstein equations (\ref{3n}) and
(\ref{3nm}) and the equation of motions (\ref{5n}), which have to be
matched by singularity in the second derivatives in $y$ on the
left-hand side. Since all fields under consideration are symmetric
under the orbifold symmetry $Z_2$, these jumps in the first
derivatives in $y$ fix these first derivatives completely at $y=0$.
Here, these junction conditions read
\begin{eqnarray}
  &&\frac{a'}{a}|_{y=0}= -\frac{k^2_5}{6}(\rho_b+\xi),\nonumber \\
  &&n'|_{y=0} = \frac{k^2_5}{6}(3p_b+2\rho_b-\xi),\label{6n}
\end{eqnarray}
and
\begin{eqnarray}
&&  \phi'|_{y=0} = \frac{1}{2}\frac{\delta \mathcal{L}_b^{mat}}{\delta \phi},\nonumber\\
  &&\sigma'|_{y=0}=-\frac{1}{2}\frac{\delta \mathcal{L}_b^{mat}}{\delta
  \sigma}.\label{6nl}
\end{eqnarray}
Using the components 00 and 55 of the Einstein equation in bulk
space one can obtain
\begin{equation}\label{6nn}
    F'=\frac{2k^2_5}{3}(\Lambda-T_0^0)a^3a'-\frac{2k^2_5}{3}T_5^0a^3\dot{a},
\end{equation}
\begin{equation}\label{6nnn}
    \dot{F}=\frac{2k^2_5}{3}(\Lambda-T_5^5)a^3\dot{a}-\frac{2k^2_5}{3}n^2T_5^0a^3a',
\end{equation}
where $F$ is a function of $t$ and $y$ defined by
\begin{equation}\label{6nm}
    F(t,y)=\frac{(\dot{a}a)^2}{n^2}-(a'a)^2+ka^2.
\end{equation}
Since the quintom field does not appear in the matter field
Lagrangian on the brane (${\mathcal{L}_b^{mat}}$), Eq.(\ref{6nl})
implies that the quintom field is independent of $y$ on the brane,
namely $\phi'|_{y=0}=\sigma'|_{y=0}=0$ \cite{Ell,carol}. Therefor
the non-vanishing components of the quintom energy momentum tensor
at the location of the brane are
\begin{eqnarray}\label{7n}
 && T^0_0=-\rho_B = -\frac{1}{2}\dot{\phi}^2+\frac{1}{2}\dot{\sigma}^2-V(\phi,\sigma), \nonumber\\
 && T^i_i=p_B = \frac{1}{2}\dot{\phi}^2-\frac{1}{2}\dot{\sigma}^2-V(\phi,\sigma) ,\\
  &&T_{05}=\dot{\phi}\phi'-\dot{\sigma}\sigma'=0, \nonumber\\
  &&T^5_5= \frac{1}{2}\dot{\phi}^2-\frac{1}{2}\dot{\sigma}^2-V(\phi,\sigma),\nonumber
\end{eqnarray}
As one can see, in the case of y independent bulk quintom field
$T_{05}$ vanishes. It means that there is no flow of matter along
the fifth dimension. Using (\ref{7n}) one can solve Eq.(\ref{6nn})
which leads to the first integral of the 00 component of Einstein
equation as
\begin{equation}\label{7nm}
    \frac{k^2_5}{6}(\Lambda+\rho_B)+\frac{\mathcal{C}}{a^4}-\frac{\dot{a}^2}{a^2n^2}+\frac{a'^2}{a^2}-\frac{k}{a^2}=0,
\end{equation}
where $\mathcal{C}$ is a constant of integration which is usually
referred to dark radiation \cite{maartens}. Substituting the
junction conditions (\ref{6n}) into above equation, we arrive at the
generalized Friedmann equation on the brane as
\begin{equation}\label{8n}
 H^2+\frac{k}{a^2}=\frac{k^2_5}{6}(\Lambda+\frac{k^2_5}{6}\xi^2)+\frac{k^4_5}{18}\xi\rho_b+
   \frac{k^2_5}{6}\rho_B+\frac{k^4_5}{36}\rho_b^2+\frac{\mathcal{C}}{a^4},
\end{equation}
here $H=\frac{\dot{a}}{a}$ is the Hubble parameter. As one can see
from Eq.(\ref{8n}), in the absence of the bulk matter field, the
cosmological constant and the brane tension, the equation gives rise
to a Friedmann equation of the form $H\propto\rho_b$ instead of
$H\propto\sqrt{\rho_b}$ which is inconsistent with cosmological
observation. This problem can be solved by either considering the
cosmological constat and tension on the brane or considering a
matter field in the bulk \cite{cline,kaonti}. Recalling the junction
conditions (\ref{6n}), the 05 component of Einstein equation and
field equations (\ref{5n}) on the brane take the following form
respectively
\begin{eqnarray}
    &&\dot{\rho_b}+3H(\rho_b+p_b)=0, \label{9n}\\
   && \ddot{\phi}+3\frac{\dot{a}}{a}\dot{\phi}+\frac{\partial V(\phi,\sigma)}{\partial\phi}= 0 ,\label{10n}\\
 && \ddot{\sigma}+3\frac{\dot{a}}{a}\dot{\sigma}-\frac{\partial
  V(\phi,\sigma)}{\partial\sigma}=0.\label{11n}
\end{eqnarray}
It should be noted that if scalar field $\phi$ and $\sigma$ satisfy
the field equations (\ref{10n}) and (\ref{11n}) respectively, the
bulk energy momentum tensor is automatically conserved and we have
\begin{equation}\label{9nn}
    \dot{\rho_B}+3H(\rho_B+p_B)=0.
\end{equation}
We are interested in studying the acceleration condition for a
universe with the quintom field in the bulk. The condition for
acceleration can be obtained from (\ref{8n}) by using the
conservation equation of the brane and bulk matter field (\ref{9n})
and (\ref{9nn})
\begin{equation}\label{12n}
 \frac{\ddot{a}}{a}=\frac{k^2_5}{6}(\Lambda+\frac{k^2_5}{6}\xi^2)-\frac{k^4_5}{36}\xi(\rho_b+3p_b)-
   \frac{k^2_5}{12}(\rho_B+3p_B)-\frac{k^4_5}{36}(2\rho_b^2+3\rho_bp_b)-\frac{\mathcal{C}}{a^4}.
\end{equation}
To study the role of the bulk quintom field in the late time
acceleration phase on the brane, we ignore the effect of tension,
brane matter, cosmological constant and dark radiation
\begin{equation}\label{13n}
 \frac{\ddot{a}}{a}=-\frac{k^2_5}{12}(\rho_B+3p_B).
\end{equation}
Thus the acceleration condition for a universe with quintom dark
energy in bulk is
\begin{equation}\label{15}
    p_B<-\frac{\rho_B}{3},\hspace{0.5cm} or\hspace{0.5cm}
    \dot{\phi}^2-\dot{\sigma}^2<V(\phi,\sigma).
\end{equation}
Now we consider the 55 component of Einstein equation at the
position of the brane which leads to the Raychaudhuri equation
\begin{equation}\label{16n}
    \frac{\ddot{a}}{a}+H^2+\frac{k}{a^2}=\frac{k^2_5}{3}(\Lambda+\frac{k^2_5}{6}\xi^2)-\frac{k^2_5}{36}
    (\xi(3p_b-\rho_b)+\rho_b(3p_b+\rho_b))-\frac{k^2_5}{3}T^5_5.
\end{equation}
Using Eq.(\ref{8n}) one can rewrite the above equation as
\begin{equation}\label{17n}
    \frac{\ddot{a}}{a}=\frac{k^2_5}{6}(\Lambda+\frac{k^2_5}{6}\xi^2)-\frac{k^4_5}{36}\xi(\rho_b+3p_b)-
   \frac{k^2_5}{6}\rho_B-\frac{k^4_5}{36}(2\rho_b^2+3\rho_bp_b)-\frac{\mathcal{C}}{a^4}-\frac{k^4_5}{3}T^5_5.
\end{equation}
Comparing Eq.(\ref{12n}) with (\ref{17n}) provide a constraint on
the bulk energy momentum tensor
\begin{equation}\label{18n}
    (3p_B-\rho_B)=4T^5_5,
\end{equation}
which for the quintom field with the energy momentum tensor
(\ref{7n}) leads to
\begin{equation}\label{18m}
    \dot{\phi}^2-\dot{\sigma}^2=0.
\end{equation}
It means that the time-dependent bulk quintom field influences the
brane like a time-dependent cosmological constant which can be
written in terms of the quintom potential energy,
$p_B=-\rho_B=-V(\phi)$ \cite{time}. For a particular solution of
(\ref{18m}) in which $\phi$ and $\sigma$ are both constant on the
brane, we arrive at the natural cosmological constant induced by the
time-dependent bulk quintom field.
\section{Generalized Einstein equation in the DGP braneworld}

In the DGP model, which provide a simple mechanism to modify gravity
at large distances, it is supposed that a 3-dimensional brane is
embedded in a flat 5-dimensional bulk. This model predicts that
4-dimensional Einstein gravity is a short-distance phenomenon with
deviations showing up at large distances. The transition between
four- and higher-dimensional gravitational potentials in the DGP
model arises as a consequence of the presence of both brane and bulk
Einstein terms in the action. The DGP model includes a length scale
below which the potential has usual Newtonian form and above which
the gravity becomes 5-dimensional. The cross over scale between the
4-dimensional and 5-dimensional gravity is
$r_c=\frac{k^2_5}{2\mu^2}$ in which $\mu^2=8\pi G_4$. In this
framework, existence of a higher dimensional embedding space allows
for the existence of bulk or brane matter which can certainly
influence the cosmological evolution on the brane. Now we proceed to
obtain the generalized DGP model in which both bulk cosmological
constant $\Lambda$ and brane tension $\xi$ are non-zero. We consider
the model described by the gravitational bulk-brane action
\begin{equation}\label{1n}
    S=\int d^5x\sqrt{-g}(\frac{1}{2k^2_5}{\hspace{0.3cm}R^{\hspace{-0.7cm}(5)}}{\hspace{0.4cm}-\Lambda+{\mathcal{L}_B^{mat})}+\int
    d^4x\sqrt{-q}(\frac{1}{2\mu^2}{\hspace{0.3cm}R^{\hspace{-0.7cm}(4)}}{\hspace{0.4cm}-\xi+\mathcal{L}_b^{mat})}},
\end{equation}
where ${\hspace{0.3cm}R^{\hspace{-0.7cm}(4)}}{\hspace{0.4cm}}$ is
the Ricci scalar of the induced metric $q_{\mu\nu}$. The
5-dimensional Einstein equation takes the form
\begin{equation}\label{2d}
R^{\hspace{-0.7cm}(5)}_{AB}-\frac{1}{2}g_{AB}{\hspace{0.3cm}R^{\hspace{-0.7cm}(5)}}{\hspace{0.4cm}=k^2_5(-\Lambda
g_{AB}+T_{AB}+S_{\mu\nu}\delta^\mu_A\delta^\nu_B\delta(y_b))}.
\end{equation}
Here $T_{AB}$ is the energy momentum tensor of the bulk matter,
$\Lambda$ is the cosmological constant of the bulk spacetime and the
energy momentum tensor on the brane is given by
\begin{equation}\label{18}
    S_{\mu\nu}=-\xi q_{\mu\nu}+\tau_{\mu\nu}-\mu^{-2}U_{\mu\nu},
\end{equation}
the last term is the contribution coming from the scalar curvature
of the brane with the non-vanishing components given by
\begin{eqnarray}\label{19n}
  U_{00}&=&3(H^2+k\frac{n^2}{a^2}),\nonumber\\
  & &\\
  U_{ij}&=&(\frac{a^2}{n^2}(-H^2+2H\frac{\dot{n}}{n}-2\frac{\ddot{a}}{a})-k)\gamma_{ij}.\nonumber
  \end{eqnarray}
From 00 and $ij$ components of the Einstein equation (\ref{2d}) we
find the following junction conditions which simply relate the jumps
of derivatives of the metric across the brane to the stress tensor
inside the brane
\begin{eqnarray}\label{20n}
  &&\frac{a'}{a}|_{y=0}= -\frac{k^2_5}{6}(\rho_b+\xi)+r_c(H^2+\frac{k}{a^2}),\nonumber \\
  &&n'|_{y=0}=\frac{k^2_5}{6}(3p_b+2\rho_b-\xi)+r_c(-H^2+2\frac{\ddot{a}}{a}-\frac{k}{a^2}).
\end{eqnarray}
In the following two sections we study the Einstein equation
(\ref{2d}) when the quintom field is considered as the bulk matter
field, and a quintom dark energy confined on the brane,
respectively.

\subsection{Quintom field in bulk space}
 We consider the quintom field as the bulk matter field with Lagrangian expression in Eq.(\ref{4a}). Integrating the equation of 00
component of (\ref{2d}) around $y=0$ and using junction conditions
(\ref{20n}), we arrive at the generalized (first) Friedmann equation
\begin{equation}\label{24}
    (1+\frac{\xi k^2_5}{6\mu^2})(H^2+\frac{k}{a^2})-\frac{k^2_5}{6}(\Lambda+\frac{k^2_5}{6}\xi^2)-\frac{k^4_5}{18}\xi\rho_b
    -\frac{k_5^2}{6}\rho_B+\frac{\mathcal{C}}{a^4}=\frac{k^4_5}{36\mu^4}(-\mu^2\rho_b+3(H^2+\frac{k}{a^2}))^2.
\end{equation}
The brane-Friedmann equation (\ref{8n}) can be derived from above
equation by letting $\mu$ go to infinity. We are interested to study
the effect of the quintom field on the brane. Ignoring the
cosmological constant $\Lambda$, the brane tension and the matter
field on the brane, Eq.(\ref{24}) can straightforwardly be rewritten
as

\begin{equation}\label{25}
    H^2+\frac{k}{a^2}=\frac{1}{2r^2_c}(1+\epsilon\sqrt{1-\frac{2k_5^2}{3}\rho_Br_c^2}),
\end{equation}
here $\rho_B$ is the energy density of the bulk quintom field on the
brane derived in Eq.(\ref{7n}). The two different possible
$\epsilon$ namely $\epsilon=\pm1$, correspond to two different
embeddings of the brane into the bulk spacetime \cite{gibons}. Since
the bulk quintom field satisfies the usual energy momentum
conservation law on the brane (\ref{9nn}), we have $\rho_B\propto
a^{-3(\omega+1)}$ ($\omega$ is the state parameter). Integrating
Eq.(\ref{25}) for $k=0$ and $\omega\geq-1$ where
$\dot{\phi}>\dot{\sigma}$, shows that the scale factor $a$ diverges
at late time\footnote{The integration of Eq.(\ref{25}) leads to
$I=\int{\frac{da}{a\sqrt{1+\epsilon
\sqrt{1-\frac{\beta}{a^m}}}}}=\frac{1}{\sqrt{2}r_c}\int{dt}$ in
which $m>0$ when $\omega>-1$ and $m<0$ when $\omega<-1$ and
$\beta=\frac{2k^2_5r^2_c}{3}$. The variation of $I$ against $a$ in
different value of $m$ and $\epsilon$ are plotted in Fig.1 } (See
Figure 1). Thus the energy density of the bulk matter goes to zero
at late time and reaches a regime where it is small in comparison
with $1/r^2_c$. In the case $\omega<-1$ where
$\dot{\phi}<\dot{\sigma}$, integrating Eq.(\ref{25}) indicates a
vanishing scale factor $a$ at late time, so the matter density goes
to zero and we could use the assumption $ k_5^2\rho_B\ll 1/r_c^2$.
Therefore, in the DGP model with a quintom dark energy fluid in the
bulk space, one can expand the Einstein equation (\ref{25}) under
the condition that $k_5^2\rho_B\ll1/r^2_c$ for all range of
$\omega$. At zero order and for spatially flat metric, two different
results depend on the value of $\epsilon$ can be derived.
Considering the case $\epsilon=-1$ yields
\begin{equation}\label{26}
H^2=0,
\end{equation}
which describes an asymptotically static universe. In the other case
we take $\epsilon=1$ which leads to
\begin{eqnarray}\label{27}
 H^2=\frac{1}{r_c^2}, &or& a(t)\propto \exp(\frac{t}{r_c}).
\end{eqnarray}
This provides the self-inflationary solution at late time which is
the most important aspect of the DGP model. Therefor, the late time
behavior of the universe does not alter even if we ignore the matter
field on the brane and consider a model of the universe filled with
the bulk quintom dark energy.
\begin{figure}[htp]
\centering
\includegraphics[width=0.24\textwidth,clip]{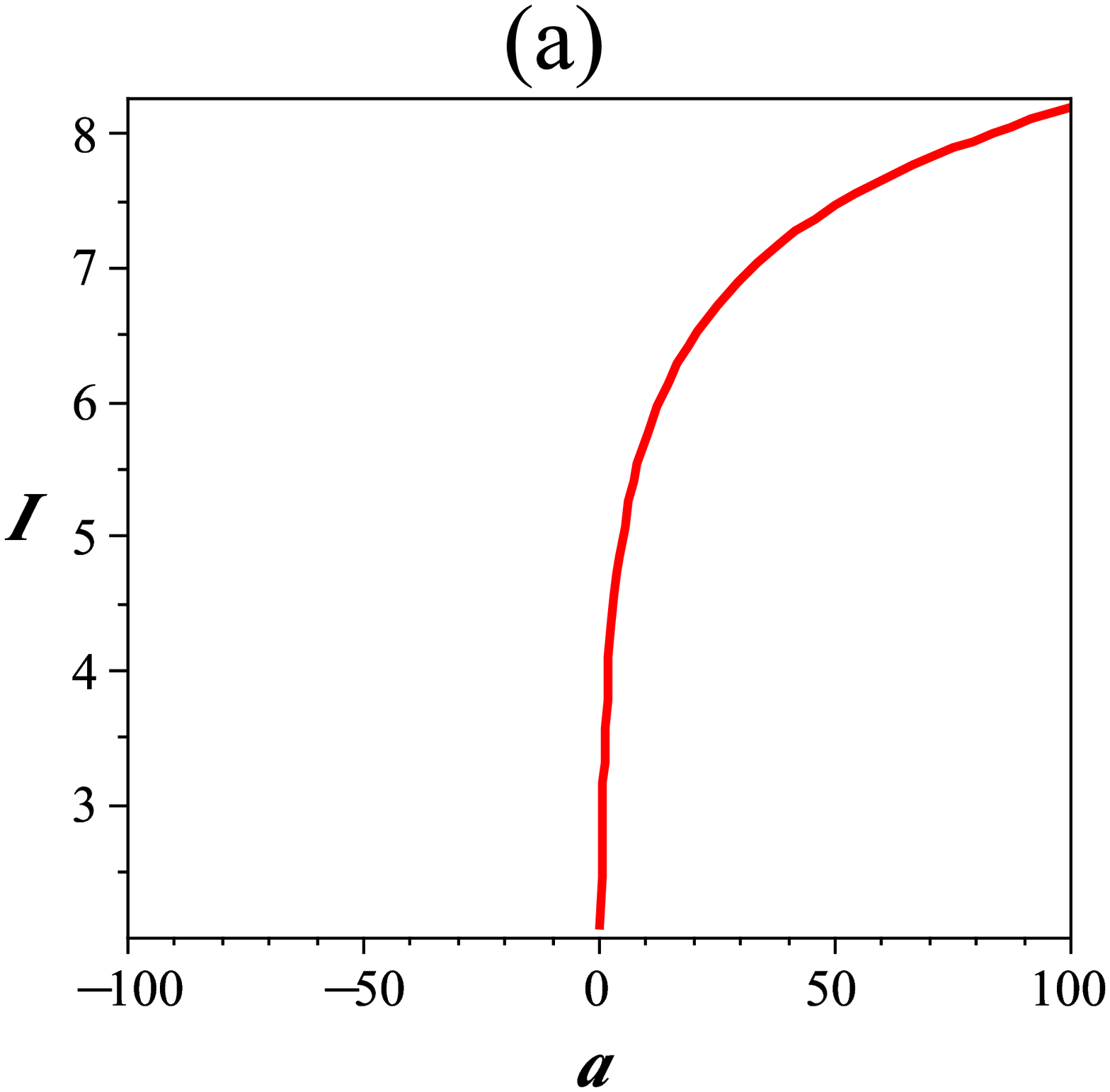}
\includegraphics[width=0.24\textwidth,clip]{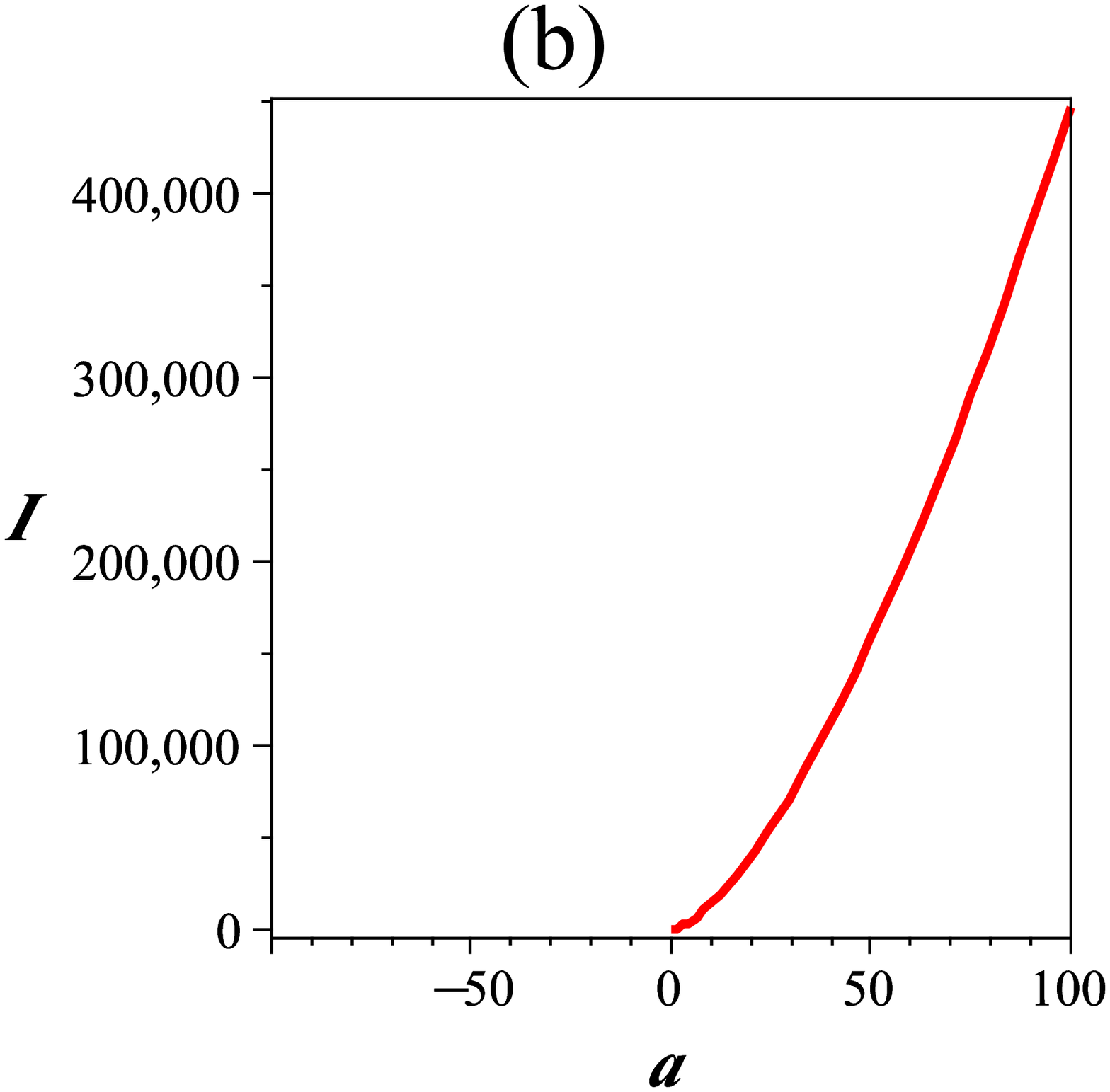}
\includegraphics[width=0.24\textwidth,clip]{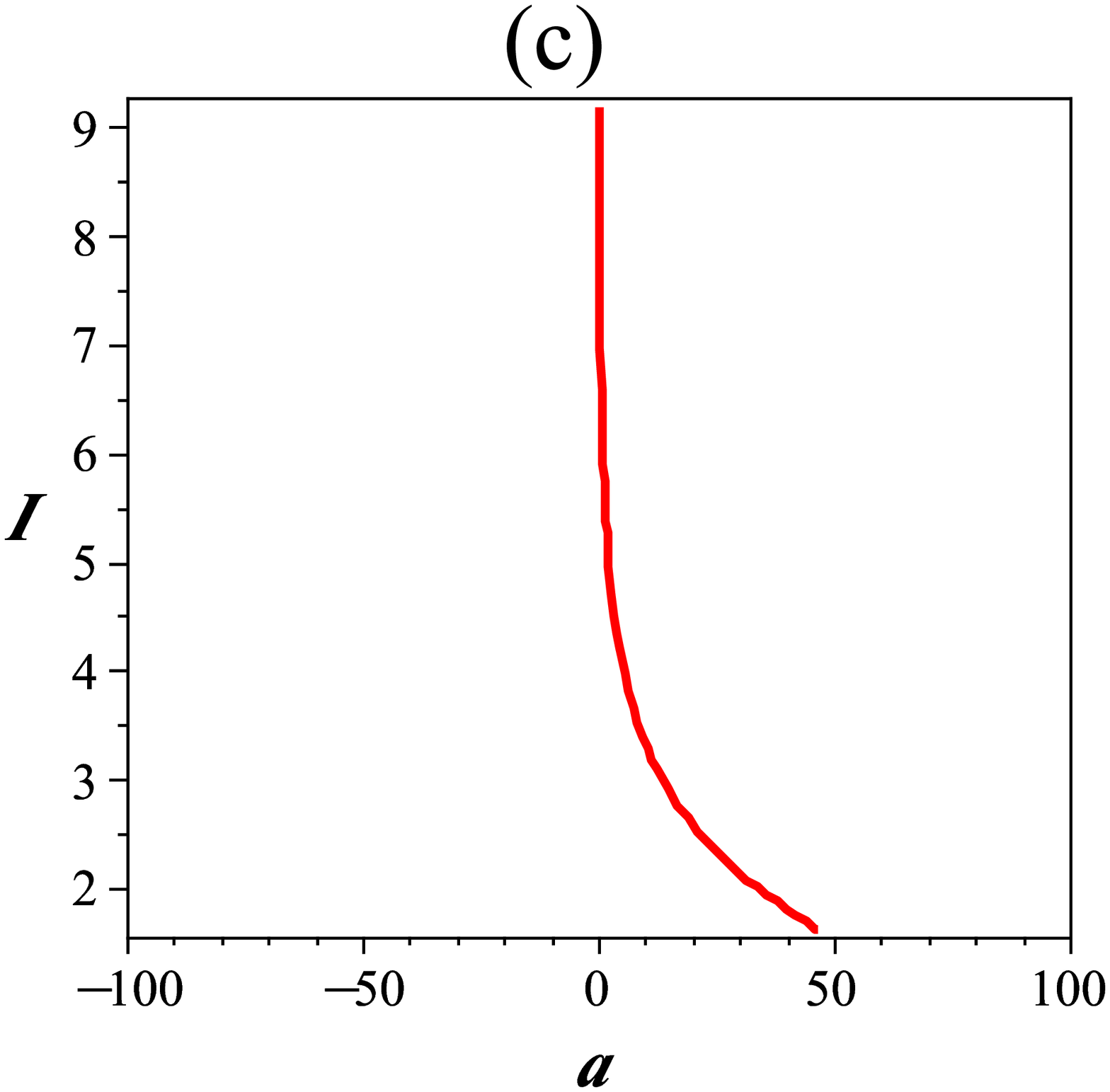}
\includegraphics[width=0.24\textwidth,clip]{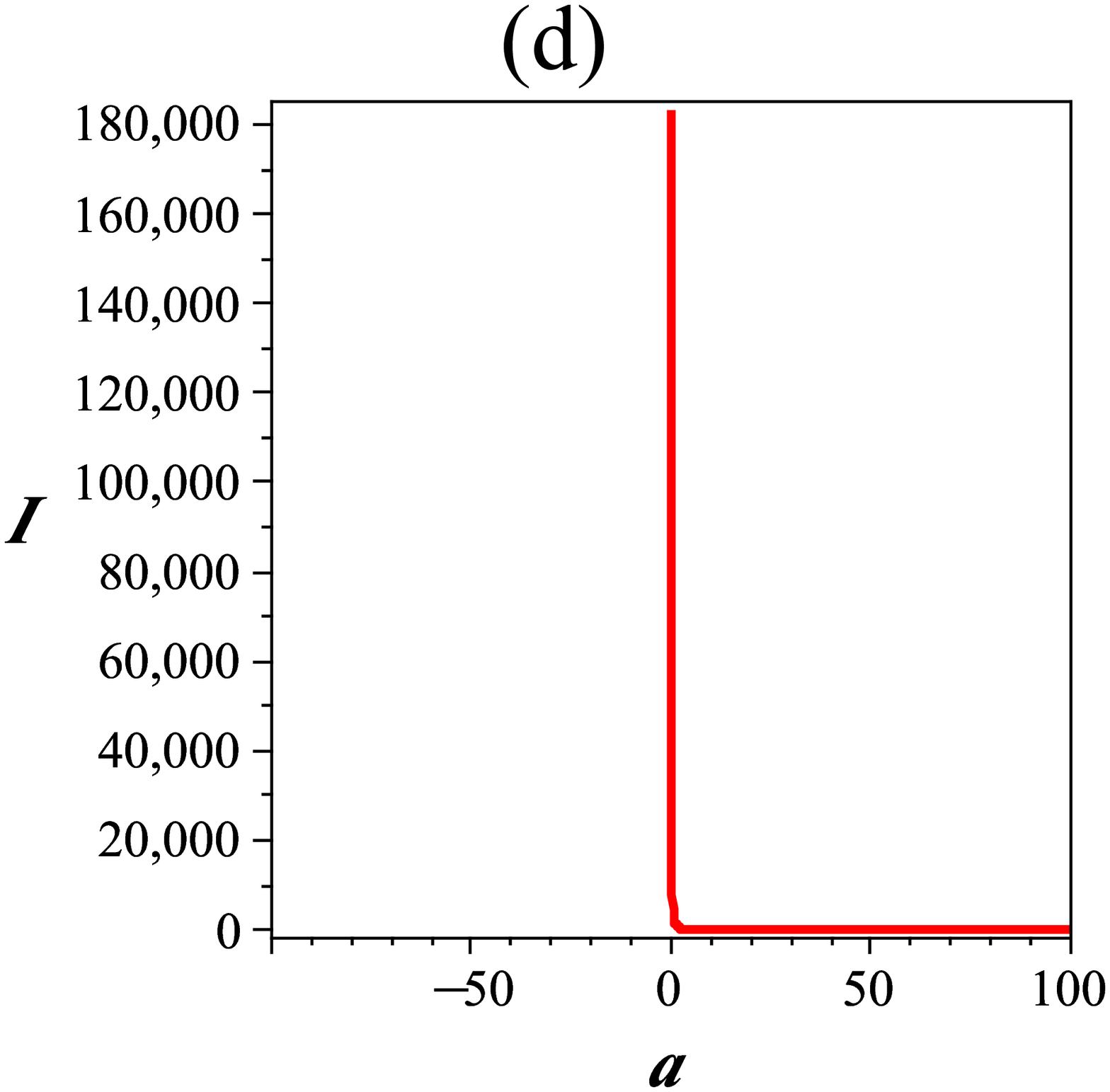}\\
\caption{These figures show the evolution of $I$  as a function of
scale factor.}
\end{figure}

\subsection{Quintom field on the brane}
In this section we ignore the bulk matter and consider a quintom
dark energy confined on the brane in the DGP model with the
Lagrangian expression as
\begin{equation}\label{1c}
    \mathcal{L}^{mat}_b=\frac{1}{2}q^{\mu\nu}(\phi_{,\mu}\phi_{,\nu}-\sigma_{,\mu}\sigma_{,\nu})
    +\tilde{V}(\phi,\sigma).
\end{equation}
The energy momentum tensor of the quintom field
 on the brane is given by
\begin{equation}\label{2c}
    \tau_{\mu\nu}=\phi_{,\mu}\phi_{,\nu}-\sigma_{,\mu}\sigma_{,\nu}-q_{\mu\nu}(\frac{1}{2}q^{\alpha\beta}
    (\phi_{,\alpha}\phi_{,\beta}
    -\sigma_{,\alpha}\sigma_{,\beta})+\tilde{V}(\phi,\sigma)).
\end{equation}
In absence of the bulk matter field and the brane tension, the
generalized Friedmann equation (\ref{24}) leads to \cite{deffayet}
\begin{equation}\label{25b}
    \sqrt{H^2+\frac{k}{a^2}}=\frac{1}{2r_c}(\epsilon+\sqrt{1+\frac{4\mu^2}{3}\rho_br^2_c}),
\end{equation}
in which $\rho_b$ is quintom energy density obtained by (\ref{2c}).
Since the energy momentum tensor on the brane is conserved, we could
apply the strategy used in the previous section to study the late
time cosmology on the brane. Integrating Eq.(\ref{25b}) for a
spatially flat spacetime yields the same result as the previous
section. It indicates that for $\omega\geq-1$ the scale factor $a$
diverges at late time \cite{deffayet} and for $\omega<-1$ the scale
factor $a$ vanishes at late time\footnote{The integration of
Eq.(\ref{25b}) leads to $I'=\int{\frac{da}{a(1+\epsilon
\sqrt{1+\frac{\beta}{a^m}})}}=\frac{1}{2r_c}\int{dt}$ in which $m>0$
when $\omega>-1$ and $m<0$ when $\omega<-1$ and
$\beta=\frac{4\mu^2r^2_c}{3}$.} (Figure $2$). Thus the energy
density of quintom dark energy goes to zero for late time and
reaches a regime where it is small in comparison with $1/r^2_c$.
Expanding the equation (\ref{25b}) under the condition
$\mu^2\rho_b\ll 1/r_c^2$ provides an asymptotically static universe,
$H=0$, in the case $\epsilon=-1$ and a self-accelerating phase,
$H=\frac{1}{r_c}$, in the case $\epsilon=1$. Therefore the presence
of quintom dark energy on the brane or in the bulk dose not change
the late time behave of the universe. In both cases for all range of
$\omega$ ($\omega<-1 $ or $\omega\geq-1$), one can derive the
self-accelerating universe at late time.

\begin{figure}[htp]
\centering
\includegraphics[width=0.24\textwidth,clip]{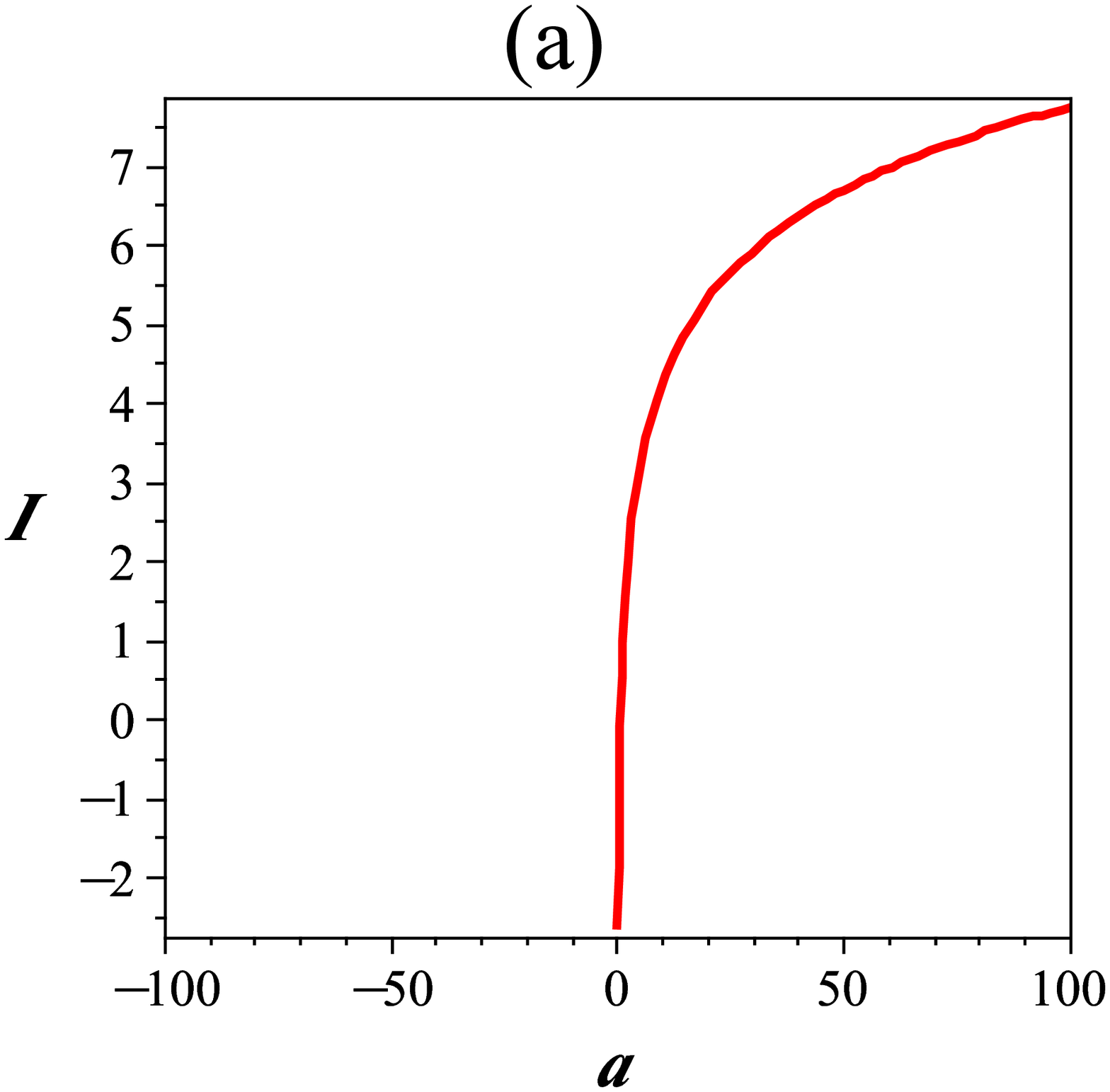}
\includegraphics[width=0.24\textwidth,clip]{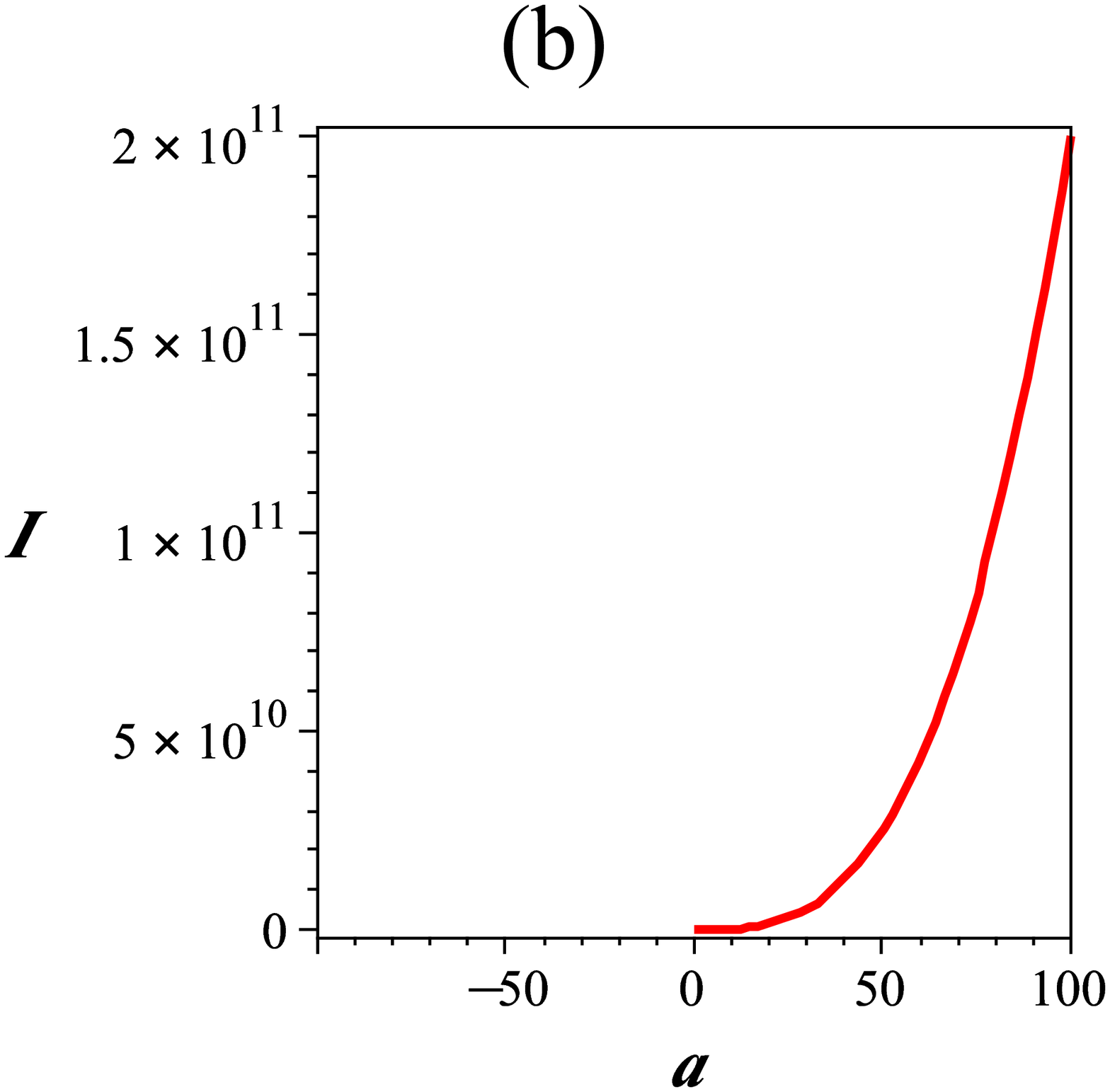}
\includegraphics[width=0.24\textwidth,clip]{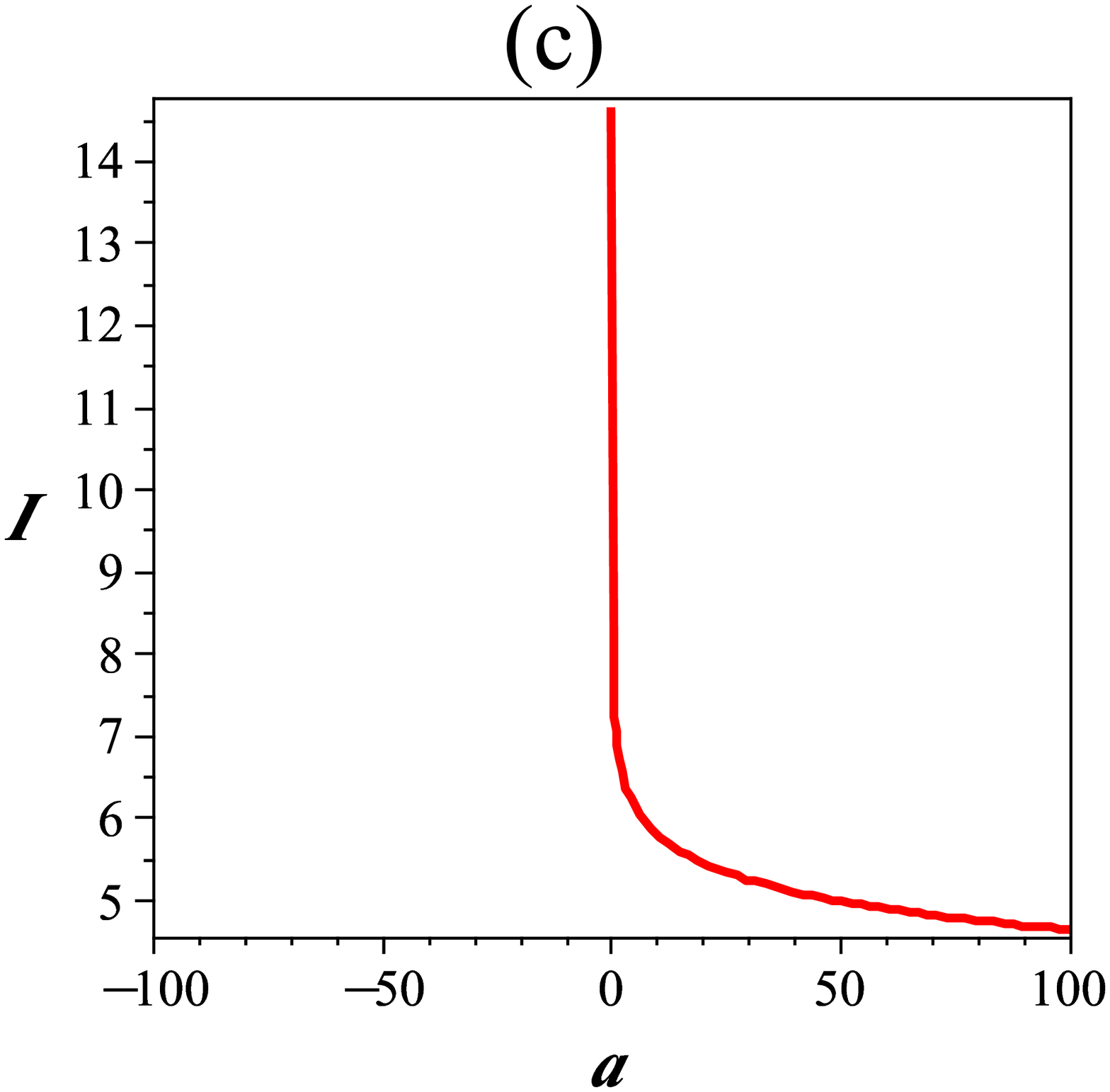}
\includegraphics[width=0.24\textwidth,clip]{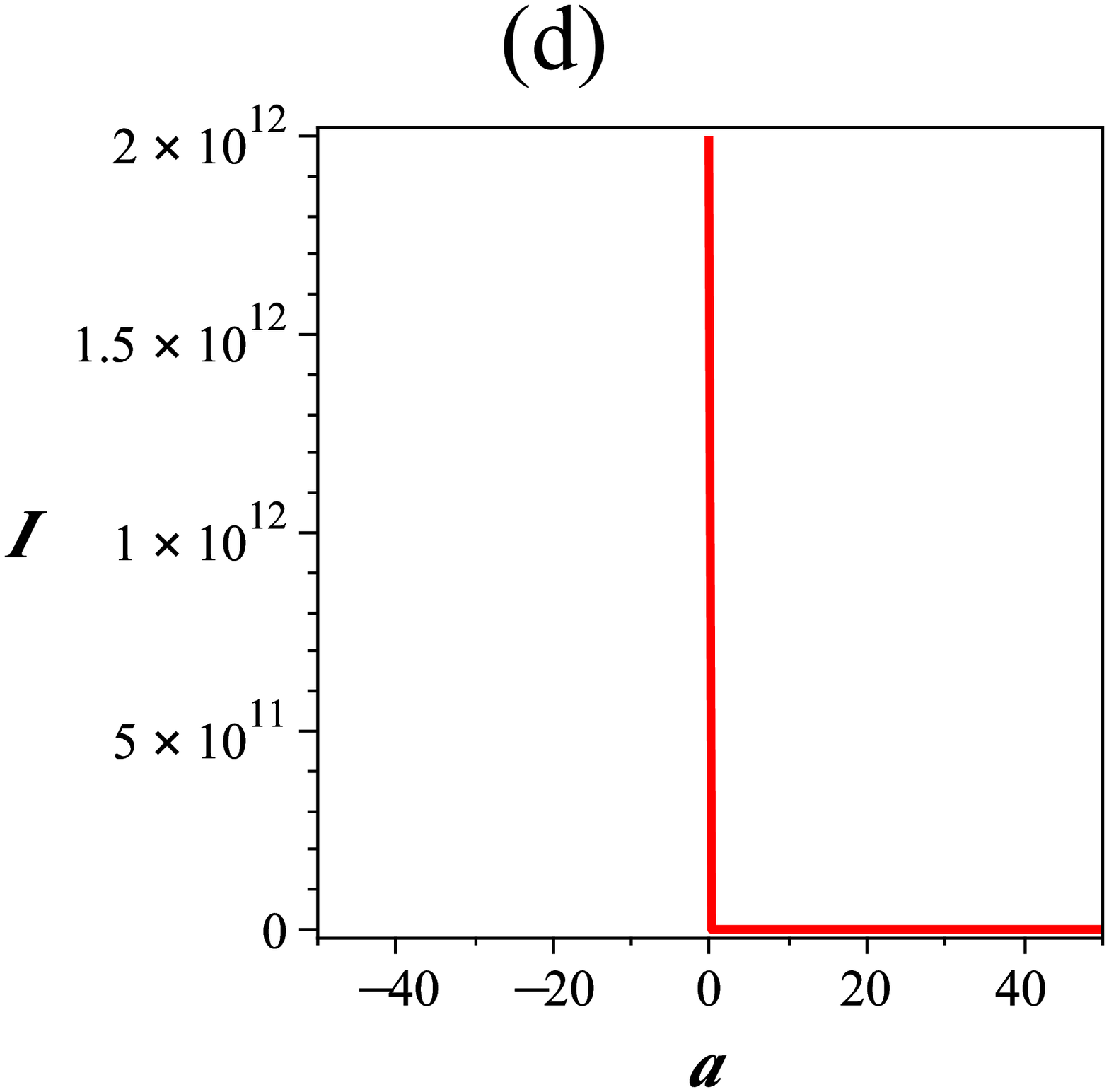}\\
\caption{These figures show the evolution of $I'$  as a function of
scale factor. }
\end{figure}

\section{Conclusion}
We have studied the cosmology of a $Z_2$ symmetrical 3-brane
embedded in a 5-dimensional spacetime including a quintom dark
energy fluid in bulk space. In the braneworld scenario, we derived
the acceleration condition due to the bulk quintom field. It was
indicated that the time-dependent bulk quintom field alters the
dynamics on the brane as a time-dependent cosmological constant
which can be derived in terms of the quintom field potential. It
means that to have an accelerated expansion phase on the brane, the
potential energy due to the bulk quintom field must be a positive
function of time.

In the DGP model, when an intrinsic curvature term is added on the
brane, we have obtained the generalized Einstein equation, where
both bulk and brane matter field are non-zero. Cosmology on 3-brane
have been studied in two different cases. In first case, we have
considered the quintom field as the bulk matter field and ignored
the brane matter. It was shown that two different solutions are
obtained for two different embeddings of the brane. In negative
branch where $\epsilon=-1$, the generalized Friedmann equation
describes an asymptotically static universe, and in positive branch
where $\epsilon=1$, we obtained $H=\frac{1}{r_c}$, which provides a
self-inflationary solution at late time. In second case, we have
considered a quintom dark energy confined on the brane in the DGP
model, and studied the solution of the generalized Friedmann
equation, when the bulk matter field and the brane tension were
ignored. Similar to the first case, here also two different results
can occur, $H=0$ for $\epsilon=-1$ and a self-inflationary solution
at late time, $H=\frac{1}{r_c}$, for $\epsilon=1$. Therefor, the
late time behavior of the universe does not alter even if we ignore
the matter field on the brane and consider a model of the universe
filled with the bulk quintom dark energy, or vice versa, we ignore
the bulk matter field and consider only the quintom dark energy on
the brane.\\
Finally we should stress on the ghost instabilities present in the
self-accelerating branch of this DGP-inspired model. The
self-accelerating branch of the DGP model contains a ghost at the
linearized level \cite{27}. Since the ghost carries negative energy
density, it leads to the instability of the spacetime. The presence
of the ghost can be attributed to the infinite volume of the
extra-dimension in DGP setup. When there are ghosts instabilities in
self-accelerating branch, it is natural to ask what are the results
of solutions decay. As a possible answer we can state that since the
normal branch solutions are ghost-free, one can think that the
self-accelerating solutions may decay into the normal branch
solutions. In fact for a given brane tension, the Hubble parameter
in the self-accelerating universe is larger than that of the normal
branch solutions. Then it is possible to have nucleation of bubbles
of the normal branch in the environment of the self-accelerating
branch solution. This is similar to the false vacuum decay in de
Sitter space. However, there are arguments against this kind of
reasoning which suggest that the self-accelerating branch does not
decay into the normal branch by forming normal branch bubbles ( see
\cite{27} for more details). It was also shown that the introduction
of Gauss-Bonnet term in the bulk does not help to overcome this
problem \cite{28}. In fact, it is still unclear what is the end
state of the ghost instability in self-accelerated branch of DGP
inspired setups. On the other hand, quintom scalar fields and
induced gravity in our setup provides a new degree of freedom which
requires special fine tuning and this may provide a suitable basis
to treat ghost instability. It seems that in our model this
additional degree of freedom has the capability to provide the
background for a more reliable solution to ghost instability due to
wider parameter space.

\section{Acknowledgment}
We would like to thank an anonymous referee for helpful comments on
the paper.

\end{document}